\documentclass[a4paper,11pt]{article}
\usepackage{graphicx}

\begin{document}
\begin{center}
{\bf \Large{Hour-glass magnetic spectrum in an insulating, hole-doped\\ antiferromagnet}}

\vspace{10pt}

{\large A. T. Boothroyd$^{1}$, P. Babkevich$^{1,2}$, D. Prabhakaran$^{1}$ \& P. G. Freeman$^{3,\dag}$}

\vspace{10pt}

{\it\large
$^1$Department of Physics, University of Oxford, Clarendon Laboratory,\\ Oxford, OX1 3PU, UK

$^2$Laboratory for Neutron Scattering, Paul Scherrer Institut,\\ CH-5232 Villigen PSI, Switzerland

$^3$Institut Laue--Langevin, BP 156, 38042 Grenoble Cedex 9, France\\
$^\dag$ Present address: Helmholtz-Zentrum f\"{u}r Materialien und Energie,\\ Hahn-Meitner Platz 1, D-14109 Berlin, Germany\\
}

\end{center}


\large

{\bf
Superconductivity in layered copper-oxide compounds emerges when charge carriers are added to antiferromagnetically-ordered CuO$_2$ layers$^1$. The carriers destroy the antiferromagnetic order, but strong spin fluctuations persist throughout the superconducting phase and are intimately linked to super-conductivity$^2$. Neutron scattering measurements of spin fluctuations in hole-doped copper oxides have revealed an unusual `hour-glass' feature in the momentum-resolved magnetic spectrum, present in a wide range of superconducting and non-superconducting materials$^{3-15}$. There is no widely-accepted explanation for this feature.
One possibility is that it derives from a pattern of alternating spin and charge stripes$^{16}$, an idea supported by measurements on stripe-ordered La$_{1.875}$Ba$_{0.125}$CuO$_4$ (ref.~15). However, many copper oxides without stripe order also exhibit an hour-glass spectrum$^{3-12}$. Here we report the observation of an hour-glass magnetic spectrum in a hole-doped antiferromagnet from outside the family of superconducting copper oxides. Our system has stripe correlations and is an insulator, which means its magnetic dynamics can conclusively be ascribed to stripes. The results provide compelling evidence that the hour-glass spectrum in the copper-oxide superconductors arises from fluctuating stripes.
}
\newpage

The term `hour-glass' describes the general structure of the magnetic spectrum as a function of energy $E$ and wavevector $\bf Q$. At low energies there is a fourfold pattern of incommensurate peaks centred on the antiferromagnetic (AF) wavevector of the parent (undoped) CuO$_2$ square lattice. The peaks disperse inwards with increasing energy until they meet at the AF wavevector, then disperse outwards again. A square-shaped intensity distribution is observed above the meeting point and is rotated by 45$^{\circ}$ with respect to that below the meeting point.
The hour-glass shape is common to many if not all hole-doped copper oxides, and is especially prominent in underdoped compositions. It is observed whether the quartet of low-energy peaks is aligned along the Cu--O bonds or at 45$^{\circ}$ to them, as found in lightly-doped La$_{2-x}$Sr$_x$CuO$_4$ (ref.~13). However, owing to the influence of superconductivity on the spin fluctuations there are some important system- and temperature-dependent variations at low energies$^{17}$.

Our experiment was performed on a member of the La$_{2-x}$Sr$_x$CoO$_4$ (LSCoO) series, which is isostructural with the ``214" copper-oxide family. The crystal structure of LSCoO contains near-perfect square lattices of Co atoms on well-separated CoO$_2$ layers (Fig.\ 1a). Unlike the copper oxides, LSCoO remains insulating over a wide range of doping$^{18}$. The parent phase
La$_{2}$CoO$_4$ exhibits$^{19}$ commensurate AF order below $T_{\rm N} = 275$\,K, similar to La$_{2}$CuO$_4$.
Substitution of Sr for La donates positive charge onto the CoO$_2$ layers. When $x > 0.3$ the AF order of the parent phase is modulated at 45$^{\circ}$ to the Co--O bonds$^{20}$. This is attributed to a self-organization of holes into arrays of charged stripes which create antiphase domain walls in the AF order. The existence of this stripe-like order is well established in the isostructural hole-doped nickelates La$_{2-x}$Sr$_x$NiO$_{4+y}$ (refs.~21--23) and in the 214 copper oxides with $x \approx 1/8$ (ref.~16).

We studied LSCoO with $x=1/3$. The ideal striped pattern of magnetic and charge order for $x=1/3$ (Fig.\ 1b) consists of diagonal lines of Co$^{3+}$, three lattice spacings apart, separating bands of AF-ordered Co$^{2+}$. The Co$^{2+}$ ions are in the high-spin $S=3/2$ state$^{19}$ while the Co$^{3+}$ are in the low-spin $S=0$ state and so do not carry a
moment$^{24}$. The diagonal modulation of the AF order gives rise to magnetic diffraction peaks at wavevectors ${\bf Q}_{\rm m} = {\bf Q}_{\rm AF} \pm (1/6 ,1/6,0)$ and equivalent positions in reciprocal space (Fig.\ 1c), where ${\bf Q}_{\rm AF} = (h+1/2, k+1/2, l)$ is the AF wavevector and $h$, $k$ and $l$ are integers. The AF order can equally well be modulated along the other diagonal giving peaks at ${\bf Q}_{\rm m} = {\bf Q}_{\rm AF} \pm (1/6,-1/6,0)$. In reality, peaks from both orthogonal domains are present with equal intensity.

Figure 2 is a composite image of the measured magnetic spectrum for wavevectors in the $(Q_x,Q_y,0)$ plane in reciprocal space. The elastic ($E=0$\,meV) scattering contains four magnetic peaks centred on the ${\bf Q}_{\rm m}$ positions. The peaks are significantly broader than the instrumental resolution and elongated in the direction perpendicular to the stripes. From the measured half-widths we determined correlation lengths of $\xi_{\parallel} \approx 10$\,\AA\ and $\xi_{\perp} \approx 6.5$\,\AA\ parallel and perpendicular to the stripes, respectively. The magnetic peaks are observed below a temperature $T_{\rm N} \approx 100$\,K (Supplementary Fig.~1).

The inelastic response in Fig.\ 2 begins at low energies with `legs' of scattering which emerge from the ${\bf Q}_{\rm m}$ positions. The legs disperse inwards with increasing energy and meet at $\sim$14\,meV. At this energy the intensity in the $(Q_x,Q_y,0)$ plane is peaked at ${\bf Q}_{\rm AF}$. The peak has an anisotropic shape with protrusions along the diagonals of the reciprocal lattice.  The intensity pattern shown at 25\,meV retains fourfold symmetry but is rotated 45$^{\circ}$ with respect to the pattern below 14\,meV.

The structure of the magnetic spectrum can be seen in more detail in Fig.\ 3 and Figs.\ 4a--d. The constant-energy scans in Fig. 3a show that the inward dispersion of the magnetic peaks at low energies is not accompanied by a corresponding outward-dispersing feature as would normally be expected for spin waves. Figures\ 3a and b show that above $\sim$20\,meV the maximum intensity disperses away from ${\bf Q}_{\rm AF}$.  Between 14\,meV and 20\,meV the intensity remains peaked at ${\bf Q}_{\rm AF}$. The peak positions in these and similar scans are plotted in Fig.\ 3c, revealing a dispersion in the shape of an hour-glass.  At higher energies the signal converges to the AF zone corners, where the scattering is strongly enhanced (Fig.\ 4d). Figure 3d is the magnetic signal at one such point and shows that the spectrum extends up to an energy of $\sim$45\,meV.

What is remarkable about these data is how closely they resemble the magnetic spectrum of the hole-doped copper oxides. Not only does the dispersion of the magnetic peaks have the characteristic hour-glass shape (Fig.\ 3c), but also there is (i) the same apparent absence of an outward-dispersing excitation branch emerging from ${\bf Q}_{\rm m}$ (Fig.\ 3a), and (ii) the same 45$^{\circ}$ rotation of the four-fold intensity pattern on crossing from below to above the waist of the hour-glass (Fig.\ 2 and Figs.\ 4a--c).

Since LSCoO is an insulator a local-moment description applies. Our analysis is based on the spin-wave spectrum of the ideal
period-3 stripe pattern, Fig.\ 1b. A two-dimensional model is sufficient because the inter-layer coupling is weak (Supplementary Fig.~2).
The strong XY-like single-ion anisotropy of Co$^{2+}$ is included in the model$^{26}$, and the two principal exchange parameters $J$ and $J'$ (Fig.~1b) were adjusted together with an intensity scale factor to fit the data.

Intensity calculations from the model are included in Figs.\ 3 and 4. All the prominent features of the data are reproduced by the simulations. In particular, a good description of the hour-glass dispersion (Fig.\ 3c) and of the intensity distribution in the $Q_x$--$Q_y$ plane (Fig.\ 4) is obtained. Moreover, the fitted values of $J$ and $J'$ are comparable with separate estimates of these parameters from LSCoO with $x=0$ and $x=0.5$, respectively$^{25,26}$. These findings are strong evidence that the basis of our model, i.e.\ the nature of the ground state and all the important interactions, is correct.

Analysis of the model reveals that the apparent inwards dispersion at low energies is caused by the combined effects of broadening and a larger intensity on the surface of the spin-wave cone nearest ${\bf Q}_{\rm AF}$ due to the high $J/J'$ ratio$^{27}$ (this also explains the absence of an hourglass feature in the nickelates --- see Supplementary Information). The point where the inward-dispersing branches meet is a saddle point formed by a maximum in the inter-stripe dispersion and a minimum in the intra-stripe dispersion. The superposition of dispersion surfaces from orthogonal stripe domains creates four intensity maxima above the saddle point which are rotated 45$^{\circ}$ with respect to the magnetic peaks below the saddle point. The enhanced intensity near the top of the spectrum is due to the XY-like anisotropy in LSCoO which creates a nearly flat dispersion at the highest energies (see Supplementary Information).

Although our model assumes static magnetic order, the results are relevant for slowly fluctuating stripes too. This is because neutron scattering is insensitive to fluctuations much slower than $\sim \hbar/\Delta E \sim 10^{-11}$\,s, where $\Delta E \sim 1$\,meV is the energy resolution. A system in which the order parameter fluctuates more slowly than this has nearly the same spectrum as one in which the correlations are static.

The observation of an hour-glass magnetic spectrum in LSCoO $(x=1/3)$ is significant because it provides an experimental demonstration that the hour-glass spectrum can arise from a system of slowly fluctuating magnetic stripes. This could have important implications for the copper-oxide superconductors. The striking resemblance between the magnetic spectra of the layered copper oxides and that found here (aside from superconductivity-induced effects) suggests that the magnetic fluctuations in the copper oxides have the same stripe-like characteristics as those which cause the hour-glass spectrum in LSCoO $(x=1/3)$, namely,(i) unidirectionally-modulated AF correlations, and (ii) a large ratio of the magnetic couplings parallel and perpendicular to the stripes.  A further requirement is a degree of broadening in order to smear out the spin-wave-like dispersion cones. Although the spins can be very well correlated in the copper oxides at low energies, experiments indicate that the spectrum broadens with increasing energy$^{11}$.

The general conditions for an hour-glass magnetic spectrum found here could in principle be satisfied by different classes of microscopic model. They do not imply that the local-moment picture that describes the cobaltates necessarily extends to the copper oxides. They do, however, impose significant constraints on any such model. It is likely that magnetic stripe correlations in the copper oxides would be accompanied by charge-stripe correlations$^{28}$. Therefore, our results lend support to interpretations of the hour-glass spectrum based on disordered stripes, and provide encouragement for theories of the copper-oxide superconductors in which the doped holes form a state with slowly fluctuating electronic nematic order$^{28,29}$.

\newpage


\newpage

\noindent{\bf Acknowledgements}

We thank Fabian Essler and Erica Carlson for helpful discussions.
This work was supported by the Engineering and Physical Sciences
Research Council of Great Britain and the Paul Scherrer Institut, Switzerland.

\vspace{10pt} \noindent{\bf Contributions}
D.P. prepared and characterised the single crystal samples. A.T.B., P.B. and P.G.F. performed the neutron scattering experiments. A.T.B. developed the theoretical model and P.B. performed the data analysis. A.T.B. wrote the manuscript.

\vspace{10pt} \noindent{\bf Competing financial interests}
The authors declare no competing financial interests.

\vspace{10pt}\noindent{\bf Author Information} Correspondence and requests for
materials should be addressed to A.T.B. (a.boothroyd@physics.ox.ac.uk).

\newpage

\begin{figure}
\begin{center}
\includegraphics
[width=8.5cm,bbllx=108,bblly=464,bburx=384, bbury=686,angle=0,clip=]
{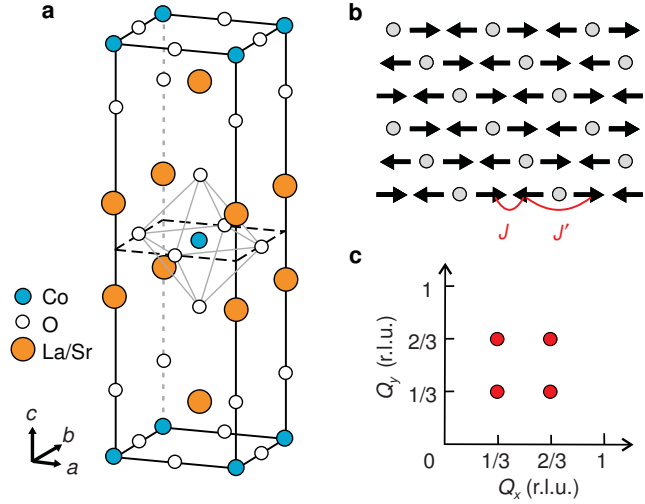} \end{center}\caption{\label{fig1} {\bf Crystal structure and magnetic order of La$_{5/3}$Sr$_{1/3}$CoO$_4$.} {\bf a}
Conventional unit cell of the tetragonal $I4/mmm$ structure. The cell parameters are $a=b=3.87$\,{\AA}, $c=12.6$\,{\AA}. {\bf b} Ideal model of magnetic and charge order in a CoO$_2$ layer. The arrows represent ordered moments on Co$^{2+}$ and the circles represent non-magnetic Co$^{3+}$. Oxygen atoms are omitted for clarity. The intra- and inter-stripe exchange interactions $J$ and $J'$ are shown. {\bf c} Diagram of reciprocal space showing the magnetic peak positions (filled circles) projected onto the $(Q_x,Q_y)$ plane. The magnetic peaks at $(1/3, 1/3)$ and $(2/3, 2/3)$ correspond to the stripe domain in {\bf b} while those at $(1/3, 2/3)$ and $(2/3, 1/3)$ are from the equivalent domain in which the stripes run along the opposite diagonal. Wavevectors are given in reciprocal lattice units (r.l.u.) relative to the tetragonal cell. }
\end{figure}

\begin{figure}
\begin{center}
\includegraphics
[width=8.5cm,bbllx=110,bblly=13,bburx=447,bbury=510,angle=0,clip=]
{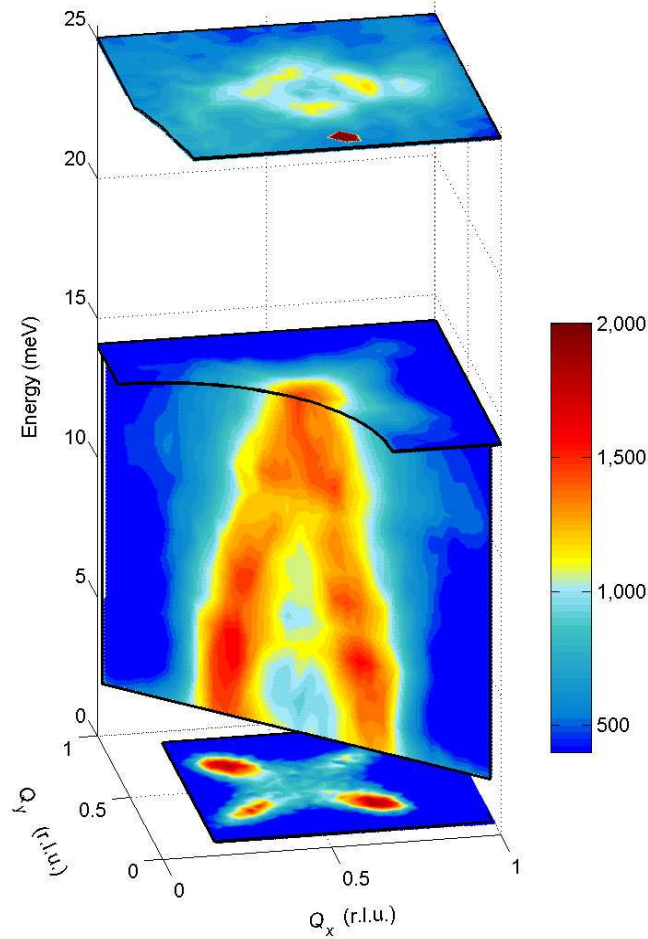} \end{center}\caption{\label{fig2} {\bf Neutron scattering intensity maps of the magnetic excitation spectrum of La$_{5/3}$Sr$_{1/3}$CoO$_4$.} The elastic and 14\,meV maps are centred on $(0.5, 0.5, 0)$, and the 25\,meV map is centred on $(1.5, 0.5, 0)$. The energy--wavevector slice was constructed from a series of constant-energy scans made at 1\,meV intervals through $(0.5, 0.5, 0)$ in the $(\xi, -\xi, 0)$ direction.  All measurements were made on a single crystal (See Supplementary Information) at a temperature of 2\,K, but a background recorded at 100\,K has been subtracted from the elastic map. Different counting times were used for each constant-energy map and so the intensity scales are not the same. The sharp feature in the 25\,meV map near $(1.5,0,0)$ is spurious. The data here and in Figs.~3 and 4 were recorded on the IN8 triple-axis spectrometer at the Institut Laue--Langevin with a fixed final energy of either 14.7\,meV or 34.8\,meV, set by Bragg reflection from a graphite analyser. The incident energy was selected by Bragg reflection from a silicon ($E<35$\,meV) or copper ($E\geq35$\,meV) monochromator. A graphite filter was placed after the sample to suppress contamination from higher orders. No collimation was used. The sample was mounted in a helium cryostat and aligned with the $a$ and $b$ axes in the horizontal scattering plane.}
\end{figure}

\begin{figure}
\begin{center}
\includegraphics
[width=8.5cm,bbllx=5,bblly=80,bburx=570, bbury=770,angle=0,clip=]
{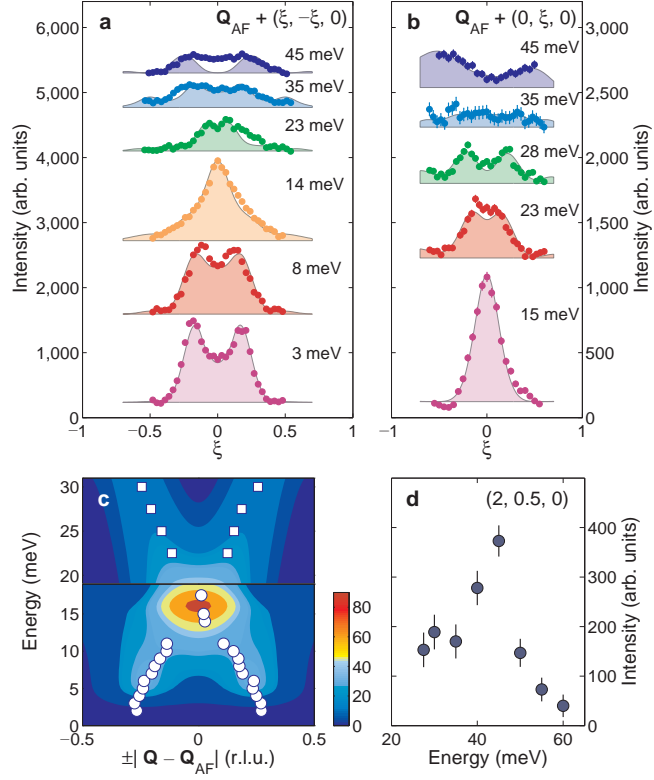} \end{center}\caption{\label{fig3} {\bf Dispersion of the magnetic excitation spectrum of La$_{5/3}$Sr$_{1/3}$CoO$_4$.} {\bf a}, {\bf b}, The variation in the scattered intensity with wavevector in diagonal and vertical scans through ${\bf Q}_{\rm AF}$ for a series of different excitation energies. We used ${\bf Q}_{\rm AF} = (0.5,0.5,0)$ for $E < 15$\,meV and ${\bf Q}_{\rm AF} = (1.5,0.5,0)$ for $E\geq 15$\,meV. Successive scans have been displaced vertically for clarity, and the intensities at 35\,meV and 45\,meV have been scaled to facilitate comparison with the lower-energy data. The shaded peaks are simulations from the spin-wave model (see Fig.~4). The intensity of the calculated signal was adjusted to match the data, and a linear background was added. {\bf c}, Dispersion of the intensity in the magnetic spectrum. The symbols represent the centres of Gaussian or Lorentzian peaks fitted to those constant-energy scans which show either two clearly resolved peaks or a single central peak, circles from scans parallel to $(\xi, -\xi,0)$ and squares from scans parallel to $(0, \xi,0)$. The intensity map is simulated from the spin-wave model. For energies below 19 meV the wavevector axis represents a diagonal scan through ${\bf Q}_{\rm AF}$, and for energies above 19\,meV it is a vertical scan. {\bf d}, Energy dependence of the magnetic intensity at the point $(2,0.5,0)$. A background measured at $(2,0,0)$ has been subtracted. Error bars are s.d. The strong peak at $\sim$45\,meV corresponds to the top of the magnetic spectrum. All measurements were made at a temperature of 2\,K. }
\end{figure}

\begin{figure}
\begin{center}
\includegraphics [width=8.5cm,bbllx=21,bblly=227,bburx=287,bbury=610,angle=0,clip=] {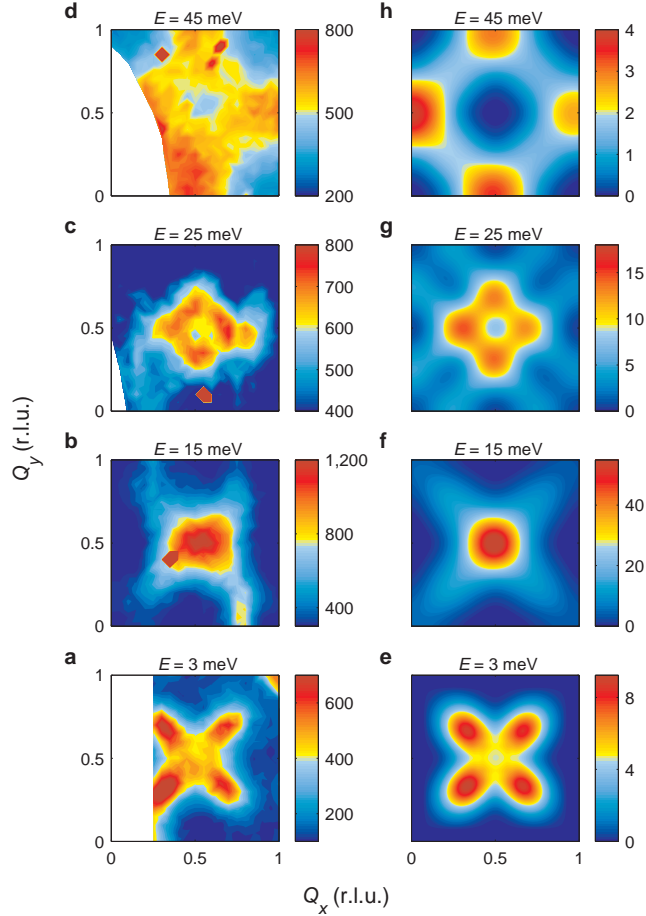} \end{center}\noindent\caption{\label{fig4}
{\bf Simulations of the magnetic spectrum of La$_{5/3}$Sr$_{1/3}$CoO$_4$.} {\bf a}--{\bf d}, Constant-energy slices through the inelastic neutron scattering data. The regions of missing data at small $\bf Q$ in the 3, 25 and 45\,meV maps are due to experimental limits on the scattering angle. Sharp spots visible on the 15, 25 and 45\,meV maps are spurious. {\bf e}--{\bf h} The corresponding simulations from the spin-wave model described in ref.~26 and in the Supplementary Information. The best-fit values of the exchange parameters (see Fig.~1b) were $J=11.5$\,meV and $J'=0.55$\,meV. The fixed parameters of the model are given in the Supplementary Information. The simulated spectra were averaged over an equal population of four equivalent magnetic domains, two wavevector domains to allow for the alternative diagonals along which the stripes may run, and two spin domains to allow the ordered moments to align along the $x$ and $y$ axes with equal probability. To simulate the observed broadening, the $\delta$-function in energy in the spin-wave cross-section was replaced by a Gaussian with a standard deviation of 1\,meV and the spectrum further broadened by convolution with a two-dimensional Gaussian in wavevector with standard deviations 0.06\,r.l.u. and 0.09\,r.l.u. parallel and perpendicular to the stripes, respectively. The model includes the $\bf Q$ variation of the dipole magnetic form factor of Co$^{2+}$. }
\end{figure}

\end{document}